\documentclass[manuscript,screen]{acmart}
\usepackage{booktabs}
\usepackage{tabularx}
\usepackage{geometry}
\usepackage{enumitem}
\usepackage{multirow}
\usepackage{multibib}
\usepackage{graphicx}
\usepackage{tikz} 
\usepackage{xcolor}

\newcommand{\mybox}[4]{
    \begin{figure}[h!]
        \centering
    \begin{tikzpicture}
        \node[anchor=text,text width=.9\columnwidth, draw, rounded corners, line width=1pt, fill=#3, inner sep=5mm] (big) {\\\footnotesize#4};
        \node[draw, rounded corners, line width=.5pt, fill=#2, anchor=west, xshift=5mm] (small) at (big.north west) {#1};
    \end{tikzpicture}
    \end{figure}
}

\newcites{Appendix}{Appendix References}

\AtBeginDocument{%
  \providecommand\BibTeX{{%
    \normalfont B\kern-0.5em{\scshape i\kern-0.25em b}\kern-0.8em\TeX}}}

\setcopyright{rightsretained}
\acmDOI{}

\acmJournal{ACMJCSS}
\acmVolume{}
\acmNumber{}
\acmArticle{}
\begin{document}
\title{GenAI Against Humanity: Nefarious Applications of Generative Artificial Intelligence and Large Language Models}
\author{Emilio Ferrara}

\email{emiliofe@usc.edu}
\orcid{1234-5678-9012}
\authornotemark[1]

\affiliation{%
  \institution{University of Southern California}
  \department{Thomas Lord Department of Computer Science}
  % \streetaddress{University Park Campus}
  % \city{LA}
  \state{CA}
  \country{USA}
  \postcode{90007}
  \city{\url{https://scholar.google.com/citations?user=0r7Syh0AAAAJ}}
}
\renewcommand{\shortauthors}{Ferrara}

\begin{abstract}
Generative Artificial Intelligence (GenAI) and Large Language Models (LLMs) are marvels of technology; celebrated for their prowess in natural language processing and multimodal content generation, they promise a transformative future. But as with all powerful tools, they come with their shadows. Picture living in a world where deepfakes are indistinguishable from reality, where synthetic identities orchestrate malicious campaigns, and where targeted misinformation or scams are crafted with unparalleled precision. Welcome to the darker side of GenAI applications.
This article is not just a journey through the meanders of potential misuse of GenAI and LLMs, but also a call to recognize the urgency of the challenges ahead. As we navigate the seas of misinformation campaigns, malicious content generation, and the eerie creation of sophisticated malware, we'll uncover the societal implications that ripple through the GenAI revolution we are witnessing. From AI-powered botnets on social media platforms to the unnerving potential of AI to generate fabricated identities, or alibis made of synthetic realities, the stakes have never been higher.
The lines between the virtual and the real worlds are blurring, and the consequences of potential GenAI's nefarious applications impact us all. This article serves both as a synthesis of rigorous research presented on the risks of GenAI and misuse of LLMs and as a thought-provoking vision of the different types of harmful GenAI applications we might encounter in the near future, and some ways we can prepare for them.
\end{abstract}

%    <concept>
%        <concept_id>10003120</concept_id>
%        <concept_desc>Human-centered computing</concept_desc>
%        <concept_significance>500</concept_significance>
%        </concept>
%    % <concept>
%    %     <concept_id>10002951.10003260</concept_id>
%    %     <concept_desc>Information systems~World Wide Web</concept_desc>
%    %     <concept_significance>300</concept_significance>
%    %     </concept>
%    <concept>
%        <concept_id>10010147.10010178</concept_id>
%        <concept_desc>Computing methodologies~Artificial intelligence</concept_desc>
%        <concept_significance>300</concept_significance>
%        </concept>
%  </ccs2012>

% \ccsdesc[500]{Human-centered computing}
% \ccsdesc[300]{Computing methodologies~Artificial intelligence}


% \received{}
% \received[revised]{}
% \received[accepted]{}

\begin{teaserfigure}
% \begin{figure}[H]
    \centering
    \includegraphics[width=.9\columnwidth, clip, trim = 100 100 100 100]{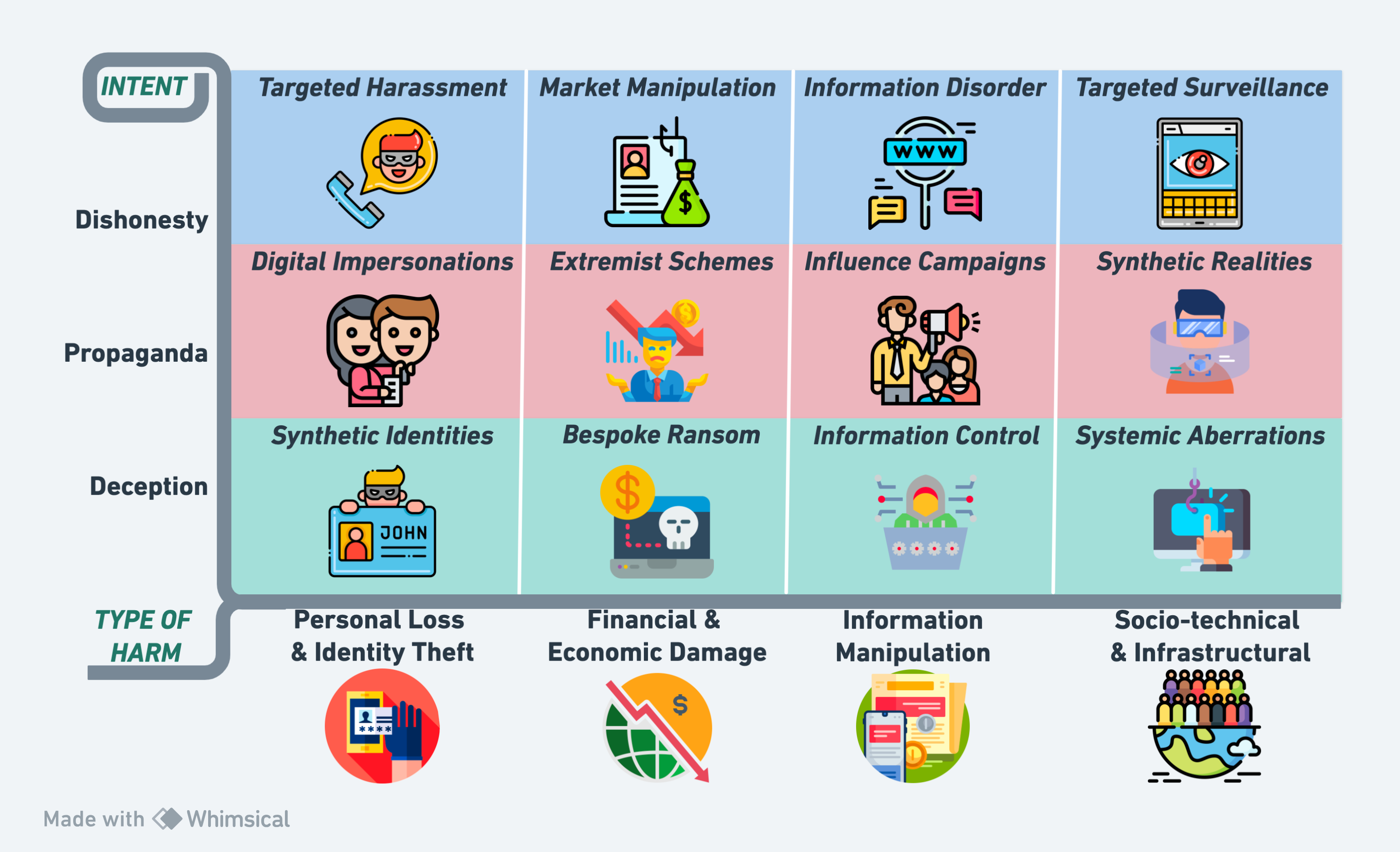}
    \caption{Charting the Landscape of Nefarious Applications of Generative Artificial Intelligence and Large Language Models}
    \label{fig:cover}
% \end{figure}
\end{teaserfigure}

\maketitle

\section*{Introduction}
In March 2019, a UK-based energy firm's CEO was duped out of \$243,000. The culprit? Not a seasoned con artist, but an AI-generated synthetic voice so convincingly mimicking the company's German parent firm's CEO that it led to a costly misstep (see Table~\ref{tab:news_articles}A).
This incident, while startling, is just the tip of the iceberg when it comes to the nefarious potential applications of Generative Artificial Intelligence (GenAI) and Large Language Models (LLMs).

GenAI and LLMs are transforming the landscape of natural language processing, content generation, and understanding. Their potential seems endless, promising innovations that could redefine the way humans and machines interact with each other or partner to work together \cite{fui2023generative, van2023chatgpt}.
However, lurking in the shadows of these advancements are challenges that threaten the very fabric of our cybersecurity, ethics, and societal structures \cite{kobis2021bad}.

This paper ventures into the darker alleys of GenAI, with a focus on LLMs. From their potential role in scaling up misinformation campaigns, to the creation of targeted scams or custom-tailor-made alibies, the risks are profound \cite{ferrara2019history}.
From the subtle perpetuation of biases to the blatant reinforcement of stereotypes \cite{schramowski2022large, caliskan2017semantics, ferrara2023should}, GenAI can become mirrors reflecting and amplifying the imperfections of our society \cite{baeza2018bias, ferrara2023butterfly}. 

Imagine a world where AI-powered botnets dominate social media \cite{ferrara2023social, yang2023anatomy}, where harmful or radicalizing content is churned out by algorithms \cite{shaw2023social}, and where the lines between reality and AI-generated content blur \cite{cao2023comprehensive}. A world where the same technology that can be used to restore lost pieces of art or ancient documents \cite{epstein2023art}, can also be used to fabricate evidence, craft alibis, and conceive the "perfect crime" (see Table~\ref{tab:news_articles}B). Many of these scenarios that until recently we would have ascribed to futuristic science fiction are already enabled by GenAI and LLMs.

As we navigate the complexities of these issues, this paper highlights the urgency of robust mitigation strategies, ethical guidelines, and continuous monitoring for GenAI and LLM systems \cite{jobin2019global, floridi2019establishing}. This exploration not only aims to summarize rigorous research on GenAI abuse, but also to ignite a discourse on the dual nature of these technologies.

\subsection*{Definition \& Mechanisms of Generative AI and LLMs}

Generative AI refers to artificial intelligence systems that can generate new content, including text, images, and audio, based on existing data \cite{cao2023comprehensive}. Unlike traditional AI, which focuses on recognizing patterns or making predictions, GenAI actively creates novel outputs. This involves complex algorithms and models that learn from large datasets, recognize underlying structures, and emulate them in unique ways \cite{epstein2023art}.
Large Language Models (LLMs), a subset of GenAI systems, specifically deal with textual data \cite{fui2023generative}. They are trained on extensive corpora of text, learning language patterns, syntax, and context. LLMs like GPT (Generative Pretrained Transformer) are capable of producing coherent, contextually relevant text, resembling human writing. Their mechanisms involve understanding input queries, accessing their extensive training data, and generating appropriate textual responses, which can range from answering questions to creating content.

\subsection*{Technological Advancements and Democratization}
The democratization of Generative AI represents a pivotal change in AI technology. The early 2020s period witnessed significant advancements in the technical capabilities of GenAI, marked by improvements in machine learning algorithms, particularly in neural networks. These advancements led to the creation of more sophisticated and efficient models that are capable of understanding and generating complex data patterns.

At the same time, there was a marked decrease in the cost of developing and deploying GenAI systems. This was due to both the falling prices of computing power and the increased availability of open-source tools and platforms, making GenAI accessible to a wider range of users and developers.
Furthermore, the proliferation of user-friendly interfaces and cloud-based services has made GenAI technologies more accessible to non-specialists. This broader access has catalyzed a wave of innovation and creativity across various sectors, allowing individuals and smaller organizations to take advantage of GenAI for various applications, thus democratizing the field of artificial intelligence.

Together, these factors have differentiated the current landscape of GenAI from previous generations of AI technologies, both in terms of technological sophistication and societal impact.

\subsection*{Regulatory Landscape}

The European Union (EU) and China have actively engaged in discussions to regulate Generative AI (GenAI). In the EU, the focus has been on establishing frameworks that ensure the ethical use of AI, focusing on data privacy, transparency, and accountability. The proposed regulations aim to categorize AI systems according to their risk levels and apply the corresponding oversight measures.
China, on the other hand, has focused on harnessing the potential of GenAI while safeguarding national security and social stability. The Chinese approach includes stringent data control measures and guidelines to prevent the misuse of AI technologies, especially in areas like surveillance and censorship.

These divergent approaches reflect the complexities and varying priorities in GenAI governance, illustrating the challenges in creating a universally accepted regulatory framework. The policies in these regions are likely to influence global standards and practices in the GenAI domain \cite{jobin2019global}.

\begin{table}[t]
\centering\footnotesize
\begin{tabular}{|c|l|l|l|}
\hline
\textbf{Ref.} & \textbf{News Title} & \textbf{Media Outlet} & \textbf{URL} \\
\hline
A & Fraudsters Used AI to Mimic CEO’s Voice in Unusual Cybercrime Case & Wall Street Journal & \href{https://www.wsj.com/articles/fraudsters-use-ai-to-mimic-ceos-voice-in-unusual-cybercrime-case-11567157402}{Link} \\
B & People Are Creating Records of Fake Historical Events Using AI & Vice & \href{https://www.vice.com/en/article/k7zqdw/people-are-creating-records-of-fake-historical-events-using-ai}{Link} \\
C & 'I don't want to upset people': Tom Cruise deepfake creator speaks out & The Guardian & \href{https://www.theguardian.com/technology/2021/mar/05/how-started-tom-cruise-deepfake-tiktok-videos}{Link} \\
D & Do These A.I.-Created Fake People Look Real to You? & New York Times & \href{https://www.nytimes.com/interactive/2020/11/21/science/artificial-intelligence-fake-people-faces.html}{Link} \\
E & Generative AI: A Blessing or a Curse for Cybersecurity? & InWeb3 & \href{https://www.inweb3.com/generative-ai-a-blessing-or-a-curse-for-cybersecurity/}{Link} \\
F & Real-world AI threats in cybersecurity aren't science fiction & VentureBeat & \href{https://venturebeat.com/ai/real-world-ai-threats-in-cybersecurity-arent-science-fiction/}{Link} \\
G & AI amplifies scam calls and other deceptions & Marketplace & \href{https://www.marketplace.org/2023/07/14/ai-amplifies-scam-calls-and-other-deceptions/}{Link} \\
H & Scammers use AI to mimic voices of loved ones in distress & CBS News & \href{https://www.cbsnews.com/news/scammers-ai-mimic-voices-loved-ones-in-distress}{Link} \\
I & Fake Or Fact? The Disturbing Future Of AI-Generated Realities & Forbes & \href{https://www.forbes.com/sites/bernardmarr/2023/07/27/fake-or-fact-the-disturbing-future-of-ai-generated-realities}{Link} \\
J & Disinformation Researchers Raise Alarms About A.I. Chatbots & New York Times & \href{https://www.nytimes.com/2023/02/08/technology/ai-chatbots-disinformation.html}{Link} \\
K & GPT-4 Produces Misinformation More Frequently, and More Persuasively, than its Predecessor & NewsGuard & \href{https://www.newsguardtech.com/misinformation-monitor/march-2023/}{Link} \\
L & The age of AI surveillance is here & Quartz & \href{https://qz.com/1060606/the-age-of-ai-surveillance-is-here}{Link} \\
M &The biggest threat of deepfakes isn’t the deepfakes themselves  & MIT Technology Review & \href{https://www.technologyreview.com/2019/10/10/132667/the-biggest-threat-of-deepfakes-isnt-the-deepfakes-themselves/}{Link} \\
N & Mushroom pickers urged to avoid foraging books on Amazon that appear to be written by AI
 & The Guardian & \href{https://www.theguardian.com/technology/2023/sep/01/mushroom-pickers-urged-to-avoid-foraging-books-on-amazon-that-appear-to-be-written-by-ai}{Link} \\
\hline
\end{tabular}
\caption{News Articles about Nefarious GenAI and LLM applications.}
\label{tab:news_articles}
\end{table}

\section*{Understanding GenAI Abuse: A taxonomy}

Figure \ref{fig:cover} offers an overview of the potential dangers associated with the misuse of generative AI models by charting the intersection between the type of harm that can be inflicted and the underlying intentions of malicious actors. 

The types of harm encompass threats to an individual's personal identity, such as identity theft, privacy breaches, or personal defamation, which we term as "Harm to the Person." Then, we have the potential for financial loss, fraud, market manipulation, and other economic harms, which fall under "Financial and Economic Damage." The distortion of the information ecosystem, including the spread of misinformation, fake news, and other forms of deceptive content \cite{vosoughi2018spread}, is categorized as "Information Manipulation." Lastly, broader harms that can impact communities, societal structures, and critical infrastructures, including threats to democratic processes, social cohesion, and technological systems, are captured under "Societal, Socio-technical, and Infrastructural Damage."

On the other side of the matrix, we have the goals (i.e., intent) of malicious actors. "Deception" involves misleading individuals or entities for various purposes, such as scams, impersonation, or other fraudulent activities \cite{kshetri2022scams}. "Propaganda" is the intent to promote a particular political, ideological, or commercial agenda, often by distorting facts or manipulating emotions. And "Dishonesty" covers a range of activities where the truth is concealed or misrepresented for personal gain, competitive advantage, or other ulterior motives. Naturally, this dimension does not fully encompass the goals or motivations behind all possible types of misuse of GenAI, but it serves as a guide to frame nefarious applications with respect to their intent to harm.

In this 3x4 matrix, each cell represents a unique combination of harm and malicious intent, illustrating the multifaceted forms of abuse possible with generative AI. For instance, AI-generated impersonation for identity theft might be found at the intersection of "Harm to the Person" and "Deception." Similarly, AI-driven fake news campaigns to influence public opinion could be represented at the crossroads of "Information Manipulation" and "Propaganda."

Table~\ref{tab:realworld} summarizes proof-of-concept examples of scenarios in which GenAI and LLMs can be intentionally misused for dishonest, propagandist, or deceiving purposes.
By understanding this framework, stakeholders can better anticipate potential threats and devise specific mitigation strategies to protect against the malicious use of generative AI.

\begin{table}[h!]
    \centering \footnotesize
    \begin{tabularx}{\textwidth}{|p{0.5cm}|p{2.2cm}|X|X|}
        \hline
        \textbf{Goal} & \textbf{Application} & \textbf{Example} & \textbf{Proof-of-Concept} \\
        \hline
        \multirow{2}{*}{\rotatebox{90}{\centering Dishonesty }} & Automated Essay Writing and Academic Dishonesty & Students could use LLMs to generate essays, research papers, or assignments, bypassing the learning process and undermining academic integrity. & Inputting a prompt like "Write a 2000-word essay on the impact of the Industrial Revolution on European society" into an LLM and receiving a detailed, well-structured essay in return. \\
        \cline{2-4}
        & Generating Fake Research Papers & LLMs can be used to produce fake research papers with fabricated data, results, and references, potentially polluting academic databases or misleading researchers. & Feeding an LLM a prompt such as "Generate a research paper on the effects of a drug called 'Zyphorin' on Alzheimer's disease" and obtaining a seemingly legitimate paper. \\
        \hline
        \multirow{3}{*}{\rotatebox{90}{\centering Propaganda }} & Impersonating Celebrities or Public Figures & LLMs can generate statements, tweets, or messages that mimic the style of celebrities or public figures, leading to misinformation or defamation. & Inputting "Generate a tweet in the style of [Celebrity Name] discussing climate change" and getting a fabricated tweet that appears genuine. \\
        \cline{2-4}
        & Automated Propaganda Generation & Governments or organizations could use LLMs to produce propaganda material at scale, targeting different demographics or regions with tailored messages. & Inputting "Generate a propaganda article promoting the benefits of a fictional government policy 'GreenFuture Initiative'" and receiving a detailed article. \\
        \cline{2-4}
        & Creating Fake Historical Documents or Texts & LLMs can be used to fabricate historical documents, letters, or texts, potentially misleading historians or altering public perception of events. & Prompting an LLM with "Generate a letter from Napoleon Bonaparte to Josephine discussing his strategies for the Battle of Waterloo" to produce a fabricated historical document. \\
        \hline
        \multirow{4}{*}{\rotatebox{90}{\centering Deception }} & Generating Fake Product Reviews & Businesses could use LLMs to generate positive reviews for their products or negative reviews for competitors, misleading consumers. & Inputting "Generate 10 positive reviews for a fictional smartphone brand 'NexaPhone'" and obtaining seemingly genuine user reviews. \\
        \cline{2-4}
        & Generating Realistic but Fake Personal Stories or Testimonies & LLMs can be used to craft personal stories or testimonies for use in deceptive marketing, false legal claims, or to manipulate public sentiment. & Inputting "Generate a personal story of someone benefiting from a fictional health supplement 'VitaBoost'" to obtain a convincing but entirely fabricated testimony. \\
        \cline{2-4}
        & Crafting Convincing Scam Emails & LLMs can be used to craft highly personalized scam emails that appear to come from legitimate sources, such as banks or service providers. & Feeding the model information about a fictional user and a prompt like "Generate an email from a bank notifying the user of suspicious account activity" to produce a scam email. \\
        \cline{2-4}
        & Crafting Legal Documents with Hidden Clauses & Unscrupulous entities could use LLMs to generate legal documents that contain hidden, misleading, or exploitative clauses. & Prompting an LLM with "Generate a rental agreement that subtly gives the landlord the right to increase rent without notice" to produce a deceptive legal document. \\
        \hline
    \end{tabularx}
    \caption{Examples of Intentional Malicious Deployments of LLMs and GenAI in the Real World.}
    \label{tab:realworld}
\end{table}

\begin{figure}[t]
    \centering
    \includegraphics[width=\columnwidth, clip, trim = 0 40 0 0]{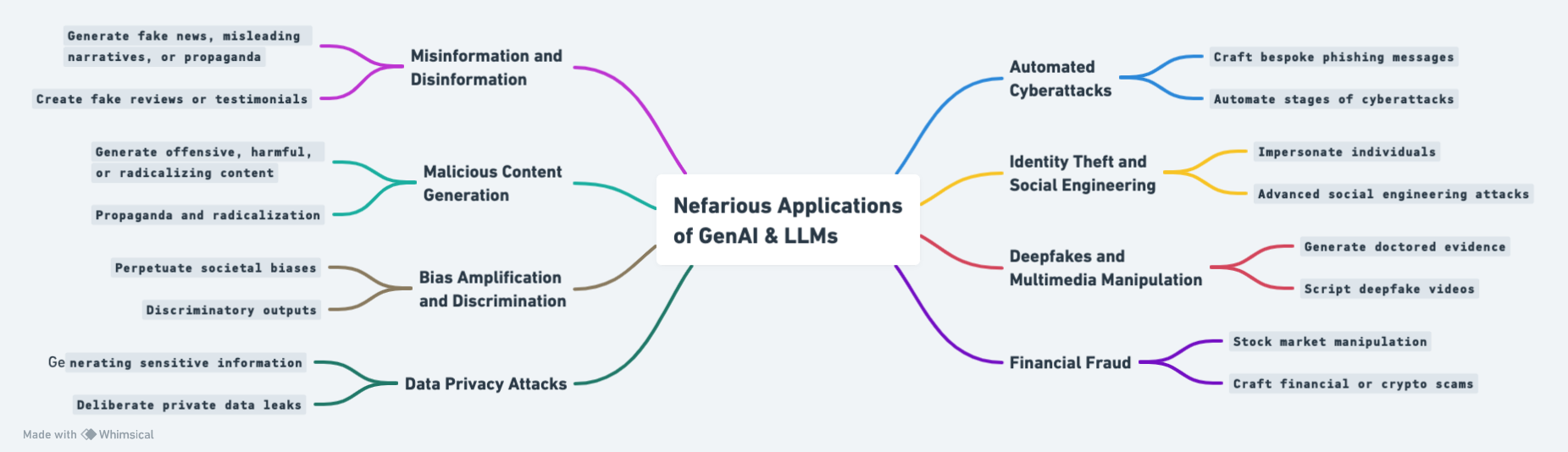}
    \caption{Mind Map of Abuse and Malicious Applications of GenAI and Large Language Models.}
    \label{fig:mindmap}
\end{figure}

\section*{A Glimpse into Days of Future Past}
Pretend for a moment that you were Tom Cruise, and on a day like any other (back in 2021) you tap into your social media feed just to see videos of yourself playing golf and prat-falling around your home (see Table~\ref{tab:news_articles}C).
What would your reaction be if you never actually recorded and posted those videos? 
The malicious use of technological advancements is barely news: each new powerful technology comes with abuse. 
The problem of tampered footage or photoshopped multimedia is not new, but GenAI and deepfake technologies have brought about a wealth of new challenges \cite{seymour2023beyond}.

The ability to create deepfakes, provide plausible deniability, and spread subliminal messages or deceiving content makes GenAI a potent tool in the hands of malicious actors. Let us unpack some of the most salient nefarious applications of GenAI technologies. Figure~\ref{fig:mindmap} provides a map of such plausible and known applications. In Table~\ref{tab:scenarios}, we summarized several proof-of-concept examples of scenarios where GenAI and LLMs can be abused to cause personal and financial harm to people, distort the information ecosystem, and manipulate sociotechnical systems and infrastructures.

\begin{table}[t]
    \centering\footnotesize
            \begin{tabularx}{\textwidth}{|p{.5cm}|p{2.2cm}|X|X|}
        \hline
        \textbf{Harm} & \textbf{Application} & \textbf{Example} & \textbf{Proof-of-Concept} \\
        \hline
        \multirow{3}{*}{\rotatebox{90}{Info. Manipulation  }} & Automated Social Media Manipulation & LLMs can be used to operate multiple social media accounts, creating an illusion of grassroots movements or artificially amplifying certain narratives. & Deploying an LLM to manage hundreds of Twitter accounts, all pushing a specific political agenda or spreading misinformation about a public health issue. \\
        \cline{2-4}
        & Generating Fake Medical Advice or Information & LLMs can produce misleading medical information, potentially endangering individuals who might act on this false advice. & Asking an LLM to "Provide natural remedies for a heart condition" and receiving potentially harmful or ineffective suggestions. \\
        \cline{2-4}
        & Crafting Deceptive Advertisements & LLMs can be used to generate advertisements that exaggerate product capabilities or make false claims. & Inputting "Create an advertisement for a fictional skincare product that provides instant results" and obtaining a misleading ad that promises unrealistic outcomes. \\
        \hline
        \multirow{2}{*}{\rotatebox{90}{Financial Harm  }} & Creating Fake Financial Reports or Data & LLMs can be used to generate false financial data or reports, potentially misleading investors or manipulating stock prices. & Prompting an LLM with "Generate a quarterly financial report for a fictional tech company 'TechNova' showing a X\% profit increase" to obtain a detailed but fabricated financial document. \\
        \cline{2-4}
        & Generating Scripts for Scam Calls & LLMs can produce scripts for scam calls, making them sound more genuine and increasing the likelihood of deceiving individuals. & Asking an LLM to "Create a script for a call claiming to be from the IRS, notifying the recipient of unpaid taxes" to produce a convincing scam script. \\
        \hline
        \multirow{3}{*}{\rotatebox{90}{Personal and Identity Harm  }} & Fake Personal Profiles and Identities & LLMs can craft detailed personal profiles, complete with background stories, for use in scams, catfishing, or espionage. & Prompting an LLM with "Generate a detailed profile of a fictional journalist named 'Alexa Morgan'" and receiving a comprehensive backstory, educational history, and career achievements. \\
        \cline{2-4}
        & Automated Online Harassment & LLMs can be deployed to target individuals online, sending them personalized and harmful messages at scale. & Using an LLM to manage multiple online accounts that continuously post derogatory comments on a specific individual's social media posts. \\
        \cline{2-4}
        & Generating Fake Evidence or Alibis & LLMs can craft detailed narratives or digital content that serve as false evidence or alibis in legal cases. & Asking an LLM to "Provide a detailed alibi for someone claiming to be at a conference in Boston from June 1-5, 2023" and receiving a comprehensive itinerary, complete with fictional events and interactions. \\
        \hline
        \multirow{2}{*}{\rotatebox{90}{Tecno-social Harm  }} & Fake Technical Support Scams & LLMs can be used to generate scripts or guides that mislead individuals into thinking they're receiving legitimate technical support, leading them to compromise their devices or data. & Prompting an LLM with "Create a guide for fixing a computer virus" and obtaining a guide that, instead, instructs users to download malicious software. \\
        \cline{2-4}
        & Generating Biased or Prejudiced Content & LLMs, if not properly fine-tuned, can produce content that reflects societal biases, potentially perpetuating stereotypes or prejudice. & Asking an LLM about descriptions of different cultures or groups and receiving outputs that contain biased or stereotypical information. \\
        \hline
        \end{tabularx}
        \caption{Proof-of-concept Scenarios Highlighting the Potential for Different Type of Harms in Malicious GenAI Applications.}
        \label{tab:scenarios}
\end{table}

% \subsection*{In a world of fakery}
% As AI models gain multi-modal capabilities, creating deepfakes---hyper-realistic forgeries of images, audio, and video---becomes easier than ever. These can be used for disinformation campaigns, fraud, and impersonation. Generative AI can also create sophisticated phishing messages or calls that can trick users into revealing sensitive information or credentials. It can even generate fake news or reviews that can harm the reputation or trust of individuals or organizations. Criminals are combining deepfakes with stolen identities to create or hijack financial accounts for illegal fund acquisition.

\subsection*{The Rise of Deepfakes}
GenAI can produce images of people that look very real, as if they could be seen on platforms like Facebook, Twitter, or Tinder. Although these individuals do not exist in reality, these synthetic identities are already being used in malicious activities (see Table~\ref{tab:news_articles}D).

\subsubsection*{AI-Generated Faces} 
There are businesses that offer "fake people" for purchase. For instance, on the website Generated.Photos, one can buy a "unique, worry-free" fake person for \$2.99 or even 1,000 people for \$1,000. If someone needs a few fake individuals, perhaps for a video game or to diversify a company website, they can obtain their photos for free from ThisPersonDoesNotExist.com. There is even a company named Rosebud.AI that can animate these fake personas and make them talk (the stated goal is for games and art, but the technology can be easily abused).

\subsubsection*{Use of Synthetic Personas} AI-generated identities are beginning to appear on the Internet and are being used by real people with malicious intentions. Examples include spies using attractive faces to infiltrate intelligence communities, right-wing propagandists hiding behind fake profiles, and online harassers using a friendly face to troll their targets.

\subsubsection*{The Perfect Alibi: Plausible Deniability and Attribution Problems}
The ability to generate fictitious images and videos can not only lend itself to abuse such as deepfake-fueled non-consensual porn generation, or the creation of misinformation for the sake of harassment or slander. Researchers are concerned that the same technologies could be used to construct alibis or fabricate criminal evidence in scalable and inexpensive ways \cite{treleaven2023future}.
Generative AI poses potential threats, especially in the realm of generating fake evidence or alibis. An article published by InWeb3 put it best in words (see Table~\ref{tab:news_articles}E):

\begin{quote}
"These possibilities undermine trust, credibility, and accountability. They create plausible deniability, the ability to deny responsibility or involvement in an action, by generating fake evidence or alibis. They also create attribution problems, the difficulty of identifying the source or origin of an action, by generating fake identities or locations. Ethical dilemmas also arise, the conflict between moral principles or values, by generating content that violates human rights or norms."   
\end{quote}

\subsection*{GenAI Against the People}
The potential threats posed by GenAI in the realm of cybersecurity include ad hominem attacks \cite{gupta2023chatgpt}, such as automated online harassment and personalized scams (see Table~\ref{tab:news_articles}F).

\subsubsection*{AI against Users} The primary targets of AI-powered attacks are not just vulnerable systems, but also human users behind those systems. AI technology can scrape personal identifiable information (PII) and gather social media data about potential victims. This enhanced data collection can help criminals craft more detailed and convincing social engineering efforts than traditional human attackers.

\subsubsection*{Bespoke Spear Phishing} While "phishing" involves generic email lures, "spear phishing" involves collecting data on a target and crafting a personalized email \cite{jagatic2007social}. Historically, spear phishing was primarily used against governments and businesses. However, with AI tools that can scrape data from various sources, spear phishing will become more common and more effective.

\subsubsection*{Automated Harassment} Beyond data theft and blackmail, GenAI can be used for automated harassment. Cybercriminals, as well as individuals with malicious intent, can use GenAI technology to launch harassment campaigns that result in service disruptions, ruined reputations, or more traditional forms of online harassment. Victims could range from businesses to private individuals or public figures. Tactics might include the creation of fake social media accounts used to spread lies or automated phone calls using voice over IP (VoIP) services. The automation of harassment processes could create a relentless and potentially untraceable campaign against victims.

\subsection*{Fake People, Real Consequences}
The use of LLMs in conjunction with other GenAI tools can bring to life synthetic personas used for scams, swindles, and other deceptions (see Table~\ref{tab:news_articles}G).

\subsubsection*{Fake users, real money scams}
GenAI can be used to scale up the generation of synthetic personal data, including fake accounts and fake transactions (see Table~\ref{tab:news_articles}G).
For example, JPMorgan Chase discovered that its acquisition of a college financial aid platform included numerous fictitious accounts. The platform was believed to contain 4.25 million customer accounts, but the bank later found that only 300,000 were legitimate. The platform vendor allegedly hired a data scientist to fabricate the majority of the accounts.
Similarly, Wells Fargo faced penalties when it was revealed that employees had opened at least 3.5 million new accounts using data from existing customers without their consent. By creating fake PINs and email addresses, funds were transferred from legitimate to fraudulent accounts.
Fake accounts have also been a problem in the social media and online retail sectors, leading to issues like spamming, fake reviews, and user-spoofing-powered fraud. For instance, PayPal disclosed that it believed 4.5 million of its accounts were not legitimate and possibly fraudulent.

\subsubsection*{Kidnapped by a bot?}
Generative AI can copy voices and likenesses, making it possible for individuals to appear as if they are saying or doing almost anything. This technology is similar to "deepfake" videos but applies to voices.

\textbf{AI-Generated Voices in Scams}: AI-generated voices are being used to enhance scams, making them more convincing (see Table~\ref{tab:news_articles}H). For instance, people have received calls from what sounds like a relative asking for money, but the voice was generated by artificial intelligence as part of a fraudulent scheme.

\textbf{Voice Spoofing and Ransom}: Threat actors can easily obtain a few seconds of someone's voice from social media or other audio sources and use generative AI to produce entire scripts of whatever they want that person to say. This has led to scams in which children appear to call their parents asking for a wire transfer for ransom (see Table~\ref{tab:news_articles}I).

\textbf{Voice Authentication}: AI can be used to bypass voice authentication systems. For example, some financial services companies allow users to download information based on voice recognition. AI can potentially be used to mimic these voices and gain unauthorized access.

\subsection*{Opening the Floodgates to Disinformation}
LLMs have the ability to craft persuasive content that can parrot false narratives and conspiracy theories, effectively and at scale (see Table~\ref{tab:news_articles}J). Some concerned researchers recently described Large Language Models like ChatGPT as \textit{weapons of mass deception} \cite{sison2023chatgpt}. It seems undeniable that the potential for GenAI and LLMs to craft fictitious, nonfactual, inaccurate, or deceiving content is unparalleled \cite{mazurczyk2023disinformation}.

\subsubsection*{LLMs and Disinformation} Soon after the launch of ChatGPT, researchers tested its ability to produce content based on questions filled with conspiracy theories and false narratives. The AI-generated content was so convincing that \textit{Gordon Crovitz}, a co-chief executive of NewsGuard (a company that tracks online misinformation), stated, "\textit{This tool is going to be the most powerful tool for spreading misinformation that has ever been on the Internet}."

\subsubsection*{ChatGPT's Capabilities} ChatGPT can produce convincing content rapidly without revealing its sources. When supplied with disinformation-loaded questions, it can generate clean variations of the content en masse within seconds. When researchers from NewsGuard asked ChatGPT to produce content based on false narratives, the AI complied about 80\% of the time (see Table~\ref{tab:news_articles}K). For instance, when asked to write from the perspective of conspiracy theorist \textit{Alex Jones} about the Parkland shooting, ChatGPT produced content that falsely claimed the mainstream media and the government used "crisis actors" to push a gun-control agenda.

\subsection*{All Systems Down}
Yet, the potential misuse of GenAI could have its most catastrophic consequences when looking at socio-technical systems and infrastructures. 
When deployed at a planetary scale, GenAI's influence extends beyond mere technological advancements: it has the potential to profoundly impact the very foundations of our economy, democracy, and infrastructure.
Targeted surveillance, censorship, and synthetic realities have been topics of concern in research community. 

\subsubsection*{Hyper-targeted Surveillance}
Enhanced by GenAI, surveillance capabilities, such as facial recognition systems, can reach unprecedented levels of accuracy. When integrated with other individual information and online data, these systems could not only recognize but also predict individual behaviors. Such advancements, while promising in the context of security, raise alarming concerns about privacy and individual rights. We may be soon be entering an age of ubiquitous GenAI-driven surveillance (see Table~\ref{tab:news_articles}L).

\subsubsection*{Total Information Control}
The intersection of GenAI with content moderation and censorship poses significant challenges to democratic values \cite{ziems2023can}. While LLMs can efficiently detect and remove harmful content from digital platforms, the potential for misuse, especially by authoritarian regimes, is concerning. The risk of suppressing dissenting voices and curating a single narrative threatens the very essence of democracy. 

\subsubsection*{Entirely Synthetic Realities}
In the era of synthetic realities--augmented reality (AR), virtual reality (VR), and the expansive metaverse--Generative Artificial Intelligence (GenAI) stands as a powerful architect. With its capability to craft intricate and indistinguishable virtual environments, GenAI has the potential to redefine our perception of reality itself. However, this transformative power is not without its pitfalls. As these synthetic realities become increasingly immersive and indistinguishable from our physical world, there lies a profound risk of manipulation. Unscrupulous entities could exploit GenAI-generated environments to influence individuals' beliefs, emotions, and behaviors. From subtly altering virtual advertisements to resonate more with individual preferences, to creating entire virtual narratives that push specific agendas or ideologies, the potential for psychological and behavioral manipulation is vast. As we embrace the wonders of synthetic realities, it becomes imperative to remain vigilant, ensuring that the line between the virtual and the real remains discernible, and that our agency within these realms is preserved.

\subsubsection*{Systemic Aberrations}
Lastly, the ability of GenAI to manipulate public opinion can have cascading effects on planetary scale systems. From influencing stock markets to swaying election outcomes, the ramifications are vast and varied. 
In conclusion, as we navigate the intricate landscape of GenAI, it is imperative to recognize its massive scale implications. While the opportunities are immense, the challenges are equally daunting. Addressing the ethical, security, and societal concerns associated with GenAI is not just a technological endeavor but a global responsibility.

\begin{table}[t]
\centering\footnotesize
\begin{tabularx}{\textwidth}{|p{2cm}|X|X|}
\hline
\textbf{Scenario} & \textbf{Opportunity} & \textbf{Danger} \\
\hline
Fabrication of Historical Artifacts & GenAI can be used to recreate or "restore" historical artifacts or paintings. & The danger lies in the potential misuse of this capability to create fake historical artifacts and sell them as genuine, misleading historians and collectors. \\
\hline
Personalized Content Generation & GenAI can curate content tailored to individual preferences, enhancing user experience on platforms like streaming services or online shopping sites. & The same technology can be exploited to create hyper-targeted misinformation or propaganda campaigns, manipulating individuals' beliefs or behaviors. \\
\hline
Voice Synthesis and Cloning & GenAI can be used to recreate voices of historical figures or digital assistive caretakers, allowing for unique educational or therapeutic experiences. & This capability can be misused to generate fake audio recordings, leading to scams, misinformation, or even potential security breaches. \\
\hline
Voice-based Services & LLMs can enhance voice-based services, providing users with natural and engaging interactions. & LLMs, when combined with voice synthesis tools, can be used for scam calls, generating scripts that sound convincing. \\
\hline
Medical Image Generation & GenAI can generate medical images for training and educational purposes, providing medical students with diverse cases without compromising patient privacy. & The technology can be exploited to fabricate medical images, leading to misdiagnoses, fraudulent research, or insurance scams. \\
\hline
VR and AR Enhancements & GenAI can enhance VR and AR experiences, making them more immersive and realistic for education, training, or entertainment. & Misuse can lead to the creation of manipulated realities that distort historical events, spread false information, or even create harmful psychological experiences. \\
\hline
Language Translation & GenAI can break down language barriers, allowing for real-time translation and fostering global communication. & It can be misused to generate misleading translations with the intent of causing misunderstandings, conflicts, or spreading fabricated narratives. \\
\hline
Automated Social Media Content & LLMs can be used to automate content generation for businesses on social media, ensuring consistent engagement and timely responses to user queries. & LLMs can be deployed to operate multiple social media accounts, creating an illusion of grassroots movements or artificially amplifying certain narratives. \\
\hline
Medical Information & LLMs can assist in providing general medical information to users, helping spread awareness about common health issues and preventive measures. & LLMs can produce misleading medical information, potentially endangering individuals who might act on this false advice. \\
\hline
Advertisements & LLMs can assist businesses in crafting engaging advertisements and detailed product descriptions. & LLMs can be used to generate deceptive advertisements that exaggerate product capabilities or make false claims. \\
\hline
Financial Reports & LLMs can assist financial analysts in generating reports, offering insights into market trends and predictions. & LLMs can be used to generate false financial data or reports, misleading investors or manipulating stock prices. \\
\hline
\end{tabularx}
\caption{Antithetic Scenarios Demonstrating the Dual Nature of GenAI's Capabilities}
\label{tab:dualnature}
\end{table}

\section*{Dual Nature Technologies: GenAI's Double-Edged Sword}

GenAI systems, with their ability to generate content, simulate voices, and even recreate historical artifacts, have opened up a plethora of opportunities across various sectors. However, with great power comes great responsibility, and the dual nature of these technologies necessitates a comprehensive understanding of their risks and benefits.
Table~\ref{tab:dualnature} illustrates a few application scenarios where a cost-benefit analysis should inform whether the opportunity created by using GenAI far outweighs the potential danger and risks that it will enable.

Take, for instance, the restoration of historical artifacts. Generative AI has shown promise in recreating or restoring damaged historical artifacts and paintings, breathing new life into our shared cultural heritage \cite{epstein2023art}. Museums and historians can leverage this technology to provide a more immersive experience for visitors, allowing them to witness history in its full glory. Yet, the same capability can be misused to fabricate fake artifacts, misleading historians and collectors, or attempting to rewrite or distort our understanding of the past. 
Similarly, scalable and cheap creation of fake identities, fabricated documentation, or fraudolent evidence, might be enabled by GenAI's ability to create seemingly legitimate documents, whose quality might match that of costly custom-made fakes (see Table~\ref{tab:figures}, left figure).
The medical field, too, is not immune to the double-edged sword of generative AI. While AI-generated medical images can provide invaluable training resources for medical students without compromising patient privacy \cite{ricci2022addressing}, the potential for fabricating medical images poses risks of misdiagnoses, fraudulent research, and insurance scams.

The realm of personalized content generation offers both promise and danger. On the one hand, GenAI-driven curation can enhance user experiences on streaming platforms, tailoring content to individual preferences, and ensuring more enjoyable and bespoke experiences. On the other hand, this personalization can be weaponized to spread misinformation or propaganda, manipulating individual beliefs and behaviors to serve malicious agendas. Take the \textit{DeepTomCruise} example from earlier: Although these particular videos were relatively harmless, the proof-of-concept highlighted the potential misuse in more sensitive areas like politics. There have been concerns that deepfakes are being used to create fake endorsements or to spread misinformation during election campaigns (see Table~\ref{tab:news_articles}M). Digital book marketplaces have been flooded by AI-generated books, many of which lack any basic fact-checking and quality assurance, sometimes providing dangerous (i.e., hallucinated \cite{ji2023survey}) information to an inattentive reader (see Table~\ref{tab:news_articles}N). GenAI could be used to depict never-occurred events with serious public diplomacy consequences (see Table~\ref{tab:figures}, center figure). Furthermore, these technologies are sufficiently advanced to implement old-school infuence strategies like the injection of subliminal messages, in a fully automated way (see Table~\ref{tab:figures}, right figure).

\begin{table}[t]
    \centering
    \begin{tabular}{ccc}
        \includegraphics[height=.25\linewidth]{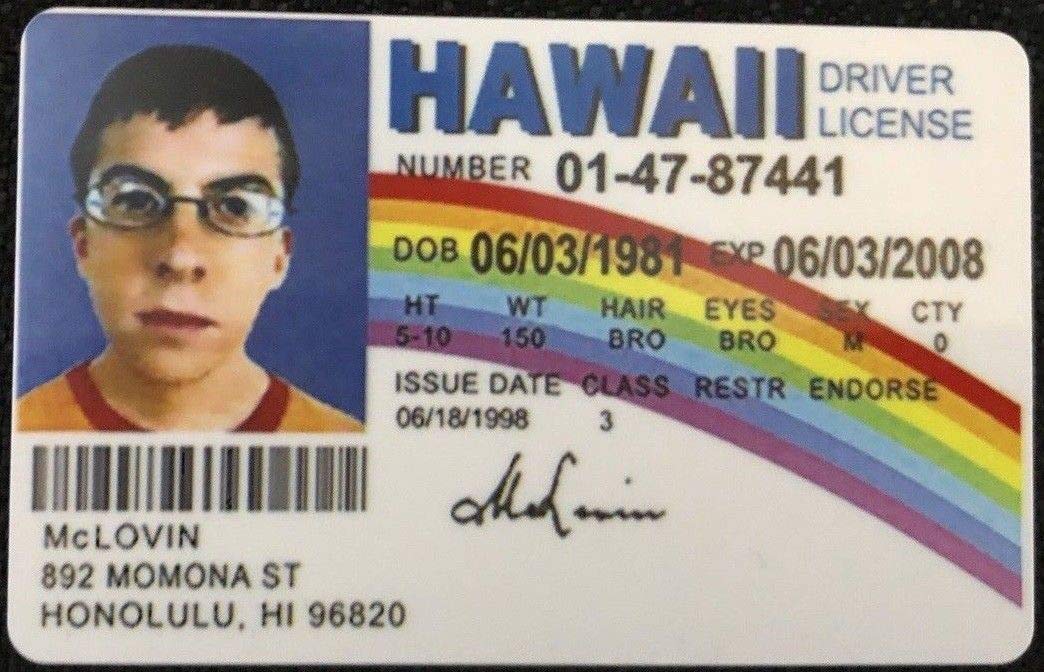} &
        \includegraphics[height=.25\linewidth]{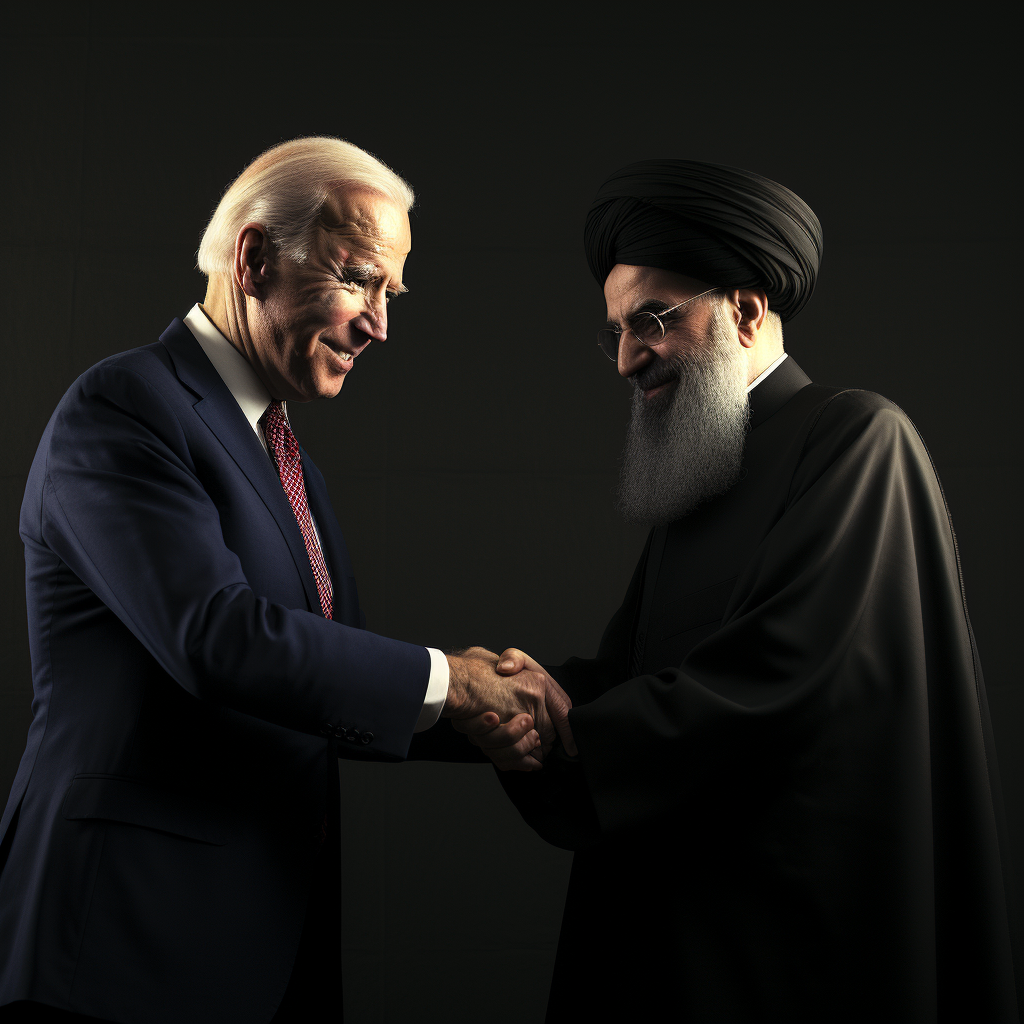}  &
        \includegraphics[height=.25\linewidth]{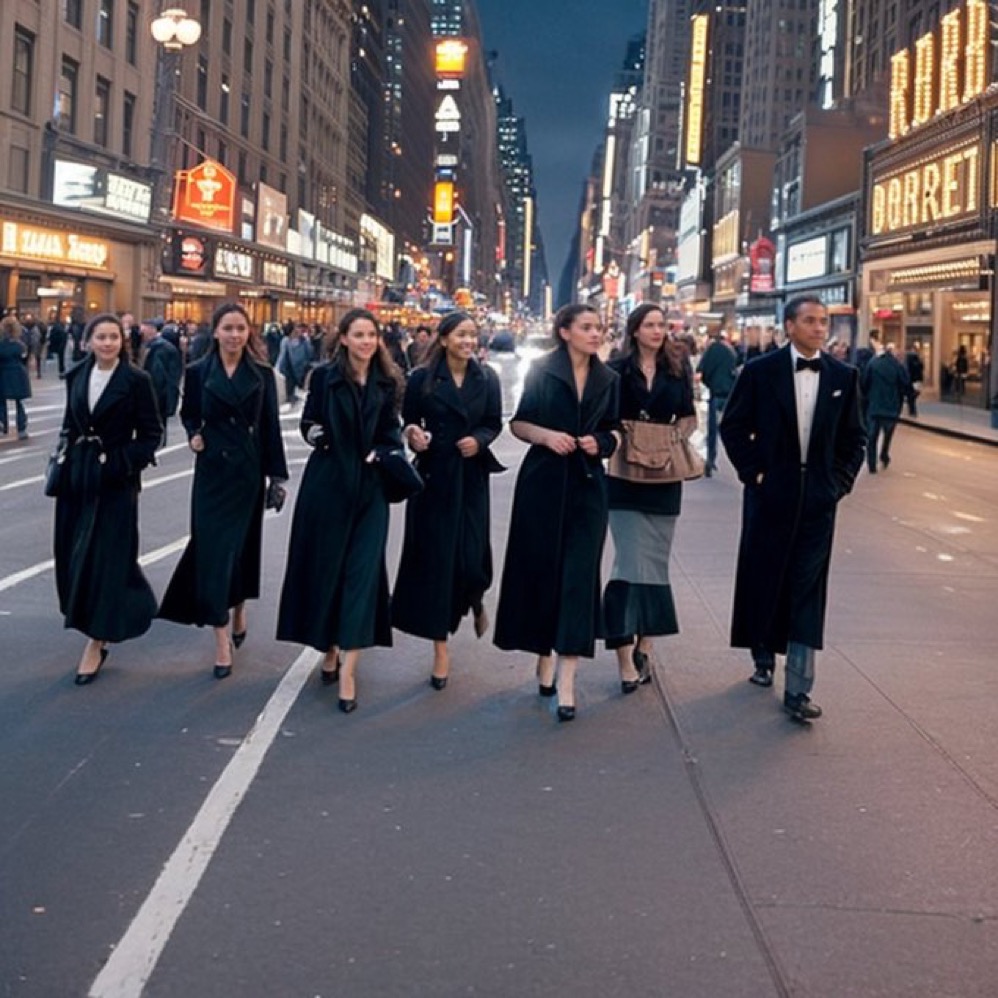} \\ 
    \end{tabular}
    \caption{(L) From fake IDs to synthetic identities, GenAI can foster a boom of fabricated documents and personas (source: \textit{Superbad} \copyright). (C) MidJourney v5 is already capable of generating lifelike depictions of never-occurred events (prompt: \textit{president biden and supreme leader of iran shaking hands}). (R) Subliminal messages can be incorporated into generated content (the optical illusion reads \textit{OBEY}).}
    \label{tab:figures}
\end{table}

\subsection*{Recommendations}
In the box "Recommendations to Mitigate GenAI Abuse", we provide a non-exhaustive list of plausible technical and socio-technical approaches that might help in mitigating GenAI abuse. 
It should be noted that most of these approaches would work for \textit{complying actors}, in other words, entities who are willing to comply with regulations, rather than having mischievous or illicit intents \cite{mozes2023use}.  
Given the inherent complexities, a robust risk mitigation strategy is imperative. This involves continuous monitoring of AI output, the establishment of ethical guidelines for the implementation of GenAI, and the promotion of transparency in AI-driven processes. Stakeholders must also be educated about the potential pitfalls of generative AI, ensuring informed decision-making at every step.

Furthermore, a comprehensive risk-benefit analysis should be conducted before deploying any generative AI system. This analysis should weigh the potential advantages against the possible harms, considering both short-term and long-term implications. Only by understanding and addressing these challenges head-on can we harness the full potential of generative AI while safeguarding our societal values and norms.

In conclusion, as we navigate the intricate tapestry of generative AI, a balanced approach that recognizes both its transformative potential and inherent risks is crucial. By adopting proactive risk mitigation strategies and conducting thorough risk-benefit analyses, we can ensure that generative AI serves as a force for good, driving innovation while preserving the integrity of our digital and physical worlds.

\mybox{\textbf{Recommendations to Mitigate GenAI Abuse}}{blue!30}{blue!10}{
\textbf{Proof of Identity}: Proof of identity refers to the verification of an individual's or entity's identity using specific documents or digital methods. Examples include technologies like humanID, OpenID, next-generation CAPTCHAs \cite{von2004telling}, etc. In GenAI:

\begin{itemize}
    \item Proof of Identity can ensure that AI-generated content or actions can be traced back to a legitimate source.
    \item Methods might include multi-factor authentication, biometric verification, or digital certificates.
\end{itemize}

\textbf{Authentication Protocols}: Authentication protocols are processes or systems used to confirm the identity of an individual, system, or entity. In the context of GenAI:

\begin{itemize}
    \item These protocols can verify whether content, actions, or requests generated by an AI system are legitimate \cite{menczer2023addressing}.
    \item Methods can include blockchain-based authentication, token-based systems, or cryptographic methods.
\end{itemize}

\textbf{Audience Disclaimers}: Audience disclaimers are explicit notifications provided to audiences to inform them about the nature of the content they are consuming.

\begin{itemize}
    \item For AI-generated content, it's crucial to inform audiences that what they're viewing, reading, or listening to was produced by an algorithm \cite{menczer2023addressing}.
    \item This promotes transparency and allows consumers to critically assess the content.
\end{itemize}

\textbf{Content Labeling}: Content labeling involves tagging content to indicate its nature, source, or other relevant attributes.

\begin{itemize}
    \item AI-generated content can be labeled to distinguish it from human-generated, ensuring users are aware of its origin.
    \item Labels can be visual tags, metadata, or even auditory cues.
\end{itemize}

\textbf{Source Verification and Provenance}: Source verification is the process of confirming the authenticity and origin of a piece of information or content.

\begin{itemize}
    \item Provenance refers to the chronology of the ownership, custody, or location of an item or piece of content.
    \item In GenAI, ensuring the provenance of data or content helps in maintaining its integrity and trustworthiness. Blockchain technology, for instance, can be used to trace the provenance of AI-generated content.
\end{itemize}

\textbf{Digital Watermarking}: Digital watermarking involves embedding a digital signal or pattern into data, making it possible to verify its authenticity or detect tampering.

\begin{itemize}
    \item For AI-generated content, watermarking can help in identifying and distinguishing it from human-generated content.
    \item It provides a layer of security and traceability, ensuring that any alterations to the original content can be detected.
\end{itemize}
}

\section*{Conclusions}
  Large Language Models (LLMs) and other Generative Artificial Intelligence (GenAI) systems have emerged as a transformative force offering unprecedented capabilities in natural language processing and multimodal content generation, and understanding. While the potential benefits of these technologies are vast, their rapid proliferation has created a myriad of malicious applications that pose significant threats to cybersecurity, ethics, and society at large. 
  
  This paper explores the darker side of generative AI applications, with a special emphasis on LLMs. We discuss potential misuse in misinformation campaigns, 
  the generation of malicious content that can bypass traditional security filters, 
  and the creation of sophisticated malware, 
  including the use of LLMs as intermediaries for malware attacks. 
  We then examine the societal implications of GenAI and LLMs, from their role in AI-powered botnets on social media to their potential to generate harmful or radicalizing content. 
  
  Our findings underscore the pressing need for robust mitigation strategies, ethical guidelines, and continuous monitoring to ensure the responsible deployment and use of GenAI and LLMs. Our aim is to raise awareness of the dual-edge nature of GenAI and LLMs, and to advocate for a balanced approach that harnesses their capabilities while safeguarding against their nefarious applications.

\begin{acks}
Work supported in part by DARPA (contract \#HR001121C0169).
\end{acks}

%%%%%%%%%%%%%%%%%%%%%%%%%%%%%%%%%%%%%%%%

\small{
\smallskip
\textbf{Data Availability Statement}: No data was used in this work. 

\textbf{COI Statement}: The author declares no conflict of interest. 
}

\bibliographystyle{ACM-Reference-Format}
\bibliography{chatgpt,genai}

%%%%%%%%%%%%%%%%%%%%%%%%%%%%%%%%%%%%%%%%%%%%%%%%%%%%%%%%%%%%%%%%
%% If your work has an appendix, this is the place to put it. %%
%%%%%%%%%%%%%%%%%%%%%%%%%%%%%%%%%%%%%%%%%%%%%%%%%%%%%%%%%%%%%%%%
% \input{sec-appendix}

\end{document}